\documentclass[twocolumn,runningheads,natbib]{svjour2}
\smartqed  
\usepackage{graphicx}
\journalname{Astrophysics and Space Science}
\begin{document}

\title{Structure of pair winds from compact objects with  
application to emission from bare strange stars}

\titlerunning{Structure of pair winds from compact objects}        

\author{
A.G. Aksenov \and M. Milgrom \and V.V. Usov
%
}


\institute{A.G. Aksenov \at
              Institute of Theoretical and Experimental Physics,
B.~Cheremushkinskaya, 25, Moscow 117218, Russia
           \and
           M. Milgrom \and V.V. Usov
           \at
           Center of Astrophysics, Weizmann Institute,
           Rehovot 76100, Israel
}

\date{Received: date / Accepted: date}

\maketitle

\begin{abstract}
We present the results of numerical simulations of stationary,
spherically outflowing, $e^\pm$ pair winds, with total
luminosities 
in the range $10^{34}- 10^{42}$ ergs~s$^{-1}$. In the concrete
example described here, the wind injection source is a hot, bare,
strange star, predicted to be a powerful source of 
$e^\pm$ pairs created by the Coulomb barrier at the quark surface.
We find that photons dominate 
in the
emerging emission, and the emerging photon spectrum is rather 
hard
and differs substantially from the thermal spectrum expected from
a neutron star with the same luminosity. This might help
distinguish the putative bare strange stars from neutron stars.
\keywords{plasmas -- radiation mechanisms: thermal -- radiative
transfer}
\PACS{52.27.Ep \and 95.30.Jx \and 97.10.Me}
\end{abstract}

\section{Introduction}
\label{intro}
There is now compelling evidence that electron-positron ($e^\pm$)
pairs form and flow away in the vicinity of many compact astronomical 
objects
(radio pulsars, accretion disk coronae of Galactic X-ray 
binaries, soft $\gamma$-ray repeaters, active galactic 
nuclei, cosmological $\gamma$-ray bursters, etc.). The
estimated luminosity in $e^\pm$ pairs varies greatly depending on
the object and the specific conditions: from $\sim
10^{31}-10^{36}$ ergs~s$^{-1}$ for radio pulsars up to
$\sim 10^{50}-10^{52}$ ergs~s$^{-1}$ in cosmological $\gamma$-ray
bursters.

For a wind out-flowing spherically from a surface of radius $R$
there is a maximum (isotropic, unbeamed) pair luminosity beyond
which the pairs annihilate significantly before they escape.
This is given by
\begin{equation}
L_\pm^{\rm max} = {4\pi m_ec^3R\Gamma^2 \over\sigma_{\rm T}}\simeq
10^{36}(R/10^6{\rm cm})\Gamma^2\,{\rm ergs~s}^{-1}, \label{Lpmmax}
\end{equation}
where $\Gamma$ is the pair bulk Lorentz factor, and $\sigma_{\rm
T}$ the Thomson cross section. When the injected pair luminosity,
$\tilde L_\pm$, greatly exceeds this value the emerging pair
luminosity , $L_\pm$ , cannot significantly exceed $L_\pm^{\rm
max}$; in this case photons strongly dominate in the emerging
emission: $L_\pm < L_\pm^{\rm max}\ll \tilde L_\pm\simeq
L_\gamma$. Injected pair luminosities typical of cosmological
$\gamma$-ray bursts (e.g., Piran 2000), $\tilde L_\pm\sim
10^{50}-10^{52}$ ergs~s$^{-1}$, greatly exceed $L_\pm^{\rm max}$. 
For such a powerful
wind the pair density near the source is very high, and the
out-flowing pairs and photons are nearly in thermal equilibrium
almost up to the wind photosphere (Paczy\'nski 1990). The
outflow process of such a wind may be described fairly well by
relativistic hydrodynamics (e.g.,  
Grimsrud \& Wasserman 1998; Iwamoto \& Takahara 2002).
\par
In contrast if  $\tilde L_\pm \ll L_\pm^{\rm max}$ annihilation of
the outflowing pairs is negligible. It is now commonly accepted
that the magnetospheres of radio pulsars contain such a very
rarefied ultra-relativistic ($\Gamma_\pm \sim 10-10^2$) pair
plasma that is practically collisionless (e.g., Melrose 1995).

Recently we developed a numerical code for solving the 
relativistic
kinetic Boltzmann equations for pairs and photons.
Using this we considered a spherically out-flowing, non-relativistic
($\Gamma\sim 1$) pair winds with the total luminosity in the
range $10^{34}-10^{42}$ ergs~s$^{-1}$, that is $\sim
(10^{-2}-10^{6})L_\pm^{\rm max}$ (Aksenov et al. 2004,
2005). (A brief account of the emerging emission from
such a pair wind has been given by Aksenov et al. 2003.)
While our numerical code can be more generally employed, 
the results presented in this paper are for
a hot, bare, strange star
as the wind injection source.
Such stars are thought to be
powerful sources of pairs created by the Coulomb barrier at the
quark surface (Usov 1998, 2001).

\begin{figure}
\centering
  \includegraphics[width=0.47\textwidth]{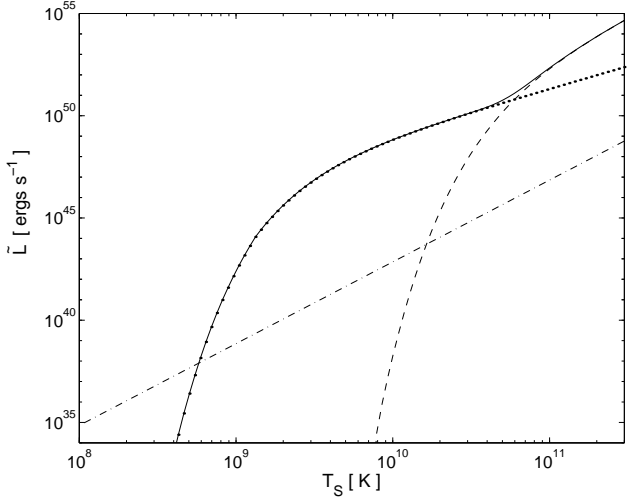}
\caption{Luminosities of a hot, bare,
strange star in $e^+e^-$ pairs (dotted line), in thermal
equilibrium photons (dashed line), and the total (solid line) as
functions of the surface temperature $T_{_{\rm S}}$. The
theoretical upper limit on the luminosity in non-equilibrium
photons, $10^{-6}L_{\rm BB}$ (Cheng \& Harko 2003), is shown by
the dot-dashed line, $L_{\rm BB}$ being the blackbody luminosity.}
\label{Aksenovf2.eps}       
\end{figure}

\section{Formulation of the problem}
We consider an $e^\pm$ pair wind that flows away from a hot, bare,
unmagnetized, non-rotating, strange star. Space-time outside the
star is described by  Schwarzschild's metric with the line element
\begin{equation}
  ds^2
 =-e^{2\phi}c^2
dt^2+e^{-2\phi}dr^2+r^2(d\vartheta^2+\sin^2\vartheta\, d\varphi^2)\,,
\label{ds1}
\end{equation}
where
\begin{equation}
  e^{\phi}=\left(1-{r_g\over r}\right)^{1/2},\,\,\,
r_g={2GM\over c^2}\simeq 2.95 \times 10^5 {M\over 
M_\odot}\,{\rm cm}\,.
\end{equation}
Following Page \& Usov (2002) we consider, as a representative
case, a strange star with a mass of $M=1.4 M_\odot$
and the circumferential radius $R =1.1\times 10^6$ cm.

The state of the plasma in the wind may be
described by the distribution functions $f_\pm ({\bf p}, r,t)$ and
$f_\gamma ({\bf p}, r,t)$ for positrons $(+)$, electrons $(-)$,
and photons, respectively, where ${\bf p}$ is the momentum of
particles. There is no emission of nuclei from the stellar
surface, so the distribution functions of positrons and electrons
are identical.

We use the general relativistic Boltzmann equations
for
the $e^\pm$ pairs and photons, whereby the distribution function
for the particles of type $i$, $f_i
(|{\bf p}|, \mu, r, t)$, ($i=e$ for
$e^\pm$ pairs and $i=\gamma$ for photons), satisfies
\begin{eqnarray}
  \frac{e^{-\phi}}{c}\frac{\partial f_i}{\partial t}
 +\frac{1}{r^2}\frac{\partial}{\partial r}(r^2\mu e^{\phi}\beta_i
f_i)
 -\frac{e^{\phi}}{p^2}\frac{\partial}{\partial p}
  \left(
    p^3 \mu \frac{\phi'}{\beta_i} f_i
  \right) \nonumber \\
 -\frac{\partial}{\partial\mu}
  \left[
    (1-\mu^2)e^\phi
          \left(\frac{\phi'}{\beta_i}-\frac{\beta_i}{r}\right) f_i
  \right]
 =\sum_q(\bar\eta^q_i-\chi^q_i f_i).
\label{dfi}
\end{eqnarray}
Here, $\mu$ is the cosine of the angle
between the radius-vector from the stellar center and
the particle momentum ${\bf p}$, $p=|{\bf p}|$,
$\beta_e=v_e/c$, $\beta_\gamma
=1$, and $v_e$ is the velocity of
electrons and positrons. Also,
$\bar\eta_i^q$ is the emission coefficient for the production of a
particle of type $i$ via the physical process labelled by $q$, and
$\chi_i^q$ is the corresponding absorption coefficient. The
summation runs over physical processes that involve a particle of
type $i$.
The processes we include are listed in
Table~1.

The thermal emission of pairs from the surface
of strange quark matter depends on the surface temperature,
$T_{\rm S}$, alone. In our simulation we use
the flux of $e^\pm$ pairs from the bare surface of a strange star
calculated by Usov (1998, 2001) as a boundary condition at the
internal computational boundary ($r=R$). Thermal emission 
of photons from the surface of a bare strange
star is strongly 
suppressed for $T_{\rm S}\ll 10^{11}$ K (Alcock et al. 1986;
Chmaj et al. 1991; Usov 2001; Cheng \& Harko 2003; Jaikumar et al. 
2004 and see Fig.~1), and we neglect this in our simulations.

The stellar surface is assumed to be a perfect mirror
for both $e^\pm$ pairs and photons.
At the external boundary 
($r=r_{\rm ext}=1.7\times 10^8$ ~cm), the pairs and photons
escape freely from the studied region.

\begin{table}
\caption{Physical Processes Included in Simulations}
\begin{center}
\begin{tabular}{ll}
  \hline \hline
  Basic Two-Body & Radiative \\
  Interaction & Variant \\
  \\ \hline
  M{\o}ller and Bhaba &  \\
  scattering & Bremsstrahlung \\
  $ee\rightarrow ee$ & $ee\leftrightarrow ee\gamma$ \\ \hline
  Compton scattering \,\,\,\,\,\,& Double Compton scattering \\
  $\gamma e\rightarrow \gamma e$ & $\gamma e\leftrightarrow \gamma
e\gamma$
  \\ \hline
  Pair annihilation & Three photon annihilation \\
  $e^+e^-\rightarrow \gamma\gamma$ & $e^+e^-\leftrightarrow
\gamma\gamma\gamma$
  \\ \hline
  Photon-photon &  \\
  pair production &   \\
  $\gamma\gamma\rightarrow e^+e^-$ &   \\ \hline
\end{tabular}
\end{center}
\end{table}

\section{Computational details}

Our grid in the $\{r, \mu,\epsilon\}$ phase-space is defined
as follows. The $r$ domain ($R<r<r_{\rm ext}$) is divided into
$j_{\rm max}$ spherical shells.
The $\mu$-grid is uniform and made 
of $k_{\rm max}$ intervals $\Delta\mu_k=2/k_\mathrm{max}$.
The energy grids for photons and electrons are both made of
$\omega_{\rm max}$ energy intervals, but the lowest energy for
photons is 0, while that for pairs is $m_ec^2$.

Here we use an $(r, \mu,\epsilon)$-grid with
$j_{\rm
max}=100$, $k_\mathrm{max}=8$, and $\omega_\mathrm{max}=13$.
The shell thicknesses are
geometrically spaced: $\Delta r_1=2\times 10^{-4}\mbox{ cm}$, and
$\Delta r_j=1.3\Delta r_{j-1}$ $(1\leq j\leq j_{\rm max})$.
The discrete energies (in keV) of the $\epsilon$-grid
minus the rest mass of the particles are 0, 2, 27, 111, 255, 353,
436, 491, 511, 530, 585, 669, 766, and $\infty$. 
A finite-difference scheme developed for solving this problem
is presented in (Aksenov et al. 2005).

\section{Numerical results}
In this section, we present the results for the structure of
the stationary $e^\pm$ winds and their emergent emission. 
Although the pair plasma ejected from the strange-star surface
contains no radiation, as the plasma moves outwards photons are
produced by pair annihilation and bremsstrahlung emission.
Figure~2 shows 
the mean optical depth for photons, from $r$ to
$r_{\rm ext}$. 
The contribution from $r_{\rm ext}$ to infinity is negligible for
$r < 10^8$ cm, so $\tau_\gamma (r)$ is practically the mean
optical depth from $r$ to infinity for these values of $r$. The
pair wind is optically thick [$\tau_\gamma (0)>1$] for $\tilde
L_\pm>10^{37}$ ergs~s$^{-1}$. The radius of the wind photosphere
$r_{\rm ph}$, determined by condition $\tau (r_{\rm ph})=1$,
varies from $\sim R$ for $\tilde L_\pm=10^{37}$ ergs~s$^{-1}$ to
$\sim 10R\simeq 10^7$ cm for $\tilde L_\pm=10^{42}$
ergs~s$^{-1}$. The wind photosphere is always deep inside our
chosen external boundary ($r_{\rm ph} < 0.1r_{\rm ext}$),
justifying our neglect of the inward $(\mu <0)$ fluxes at
$r=r_{\rm ext}$

\begin{figure}
\centering
  \includegraphics[width=0.5\textwidth]{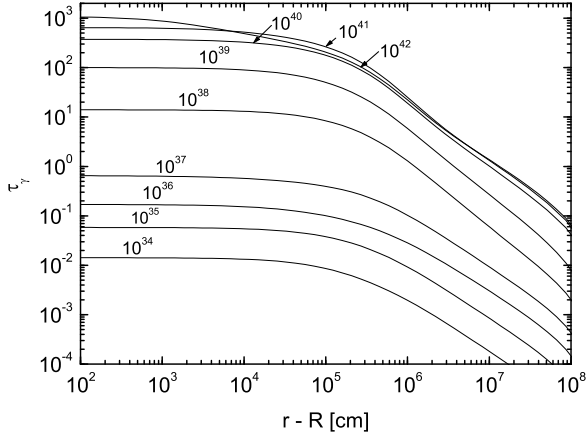}
\caption{The mean optical depth for photons, from
$r$ to $r_{\rm ext}$, as a function of the distance from the stellar
surface, for different values of $\tilde L_\pm$, as marked on the
curves.}
\label{Aksenovf2.eps}       
\end{figure}

\begin{figure}
\centering
  \includegraphics[width=0.48\textwidth]{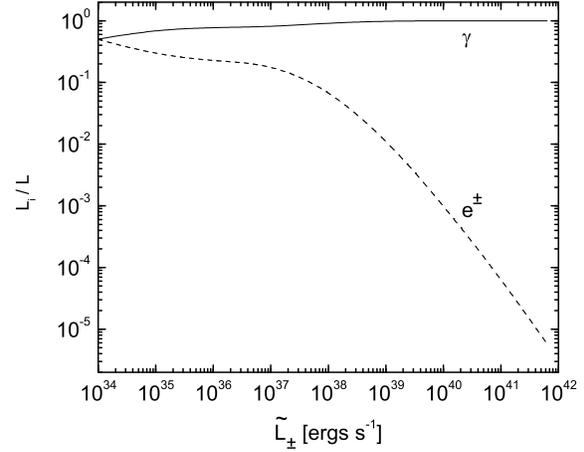}
\caption{The fractional emerging luminosities in
pairs (dashed line) and photons (solid line) as functions of the 
injected pair luminosity, $\tilde L_\pm$.}
\label{Aksenovf3.eps}       
\end{figure}

\begin{figure}
\centering
  \includegraphics[width=0.5\textwidth]{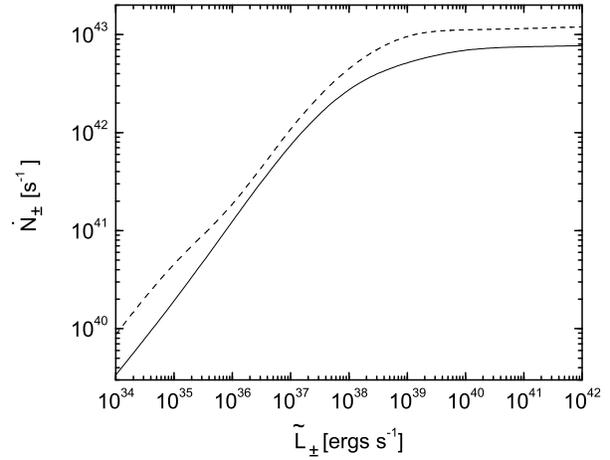}
\caption{Number rate of emerging pairs as
functions of the injected pair luminosity (solid line).
The result by Aksenov et al. (2004)
where gravity has been 
neglected is shown by
the dashed line.}
\label{Aksenovf4.eps}       
\end{figure}

Figure~3 shows the emerging luminosities in $e^\pm$ pairs ($L_e$) 
and photons
($L_\gamma$) as fractions of the total luminosity
\begin{equation}
L=L_\pm+L_\gamma =
\left(1 -{r_g\over R}\right)\tilde L_\pm\simeq 0.63 \tilde L_\pm\,.    
\end{equation}
For $L
>L_{\rm eq}\simeq 10^{34}$ ergs~s$^{-1}$, the emerging
emission consists mostly of photons ($L_\gamma >L_e$). This simply
reflects the fact that in this case the pair annihilation time
$t_{\rm ann}\sim (n_e \sigma_{\rm T}c)^{-1}$ is less than the
escape time $t_{\rm esc}\sim R/c$, so most injected pairs
annihilate before they escape. The value of $L_{\rm eq}$
is about two orders of magnitude smaller than $L^{\rm max}_\pm$
[see equation (1)] that is estimated from consideration of the same
processes, but without taking into account gravity of the star.
However, with gravity at low luminosities ($\tilde L_\pm < 10^{36}$ 
ergs~s$^{-1}$) pairs emitted by the stellar
surface are mainly captured 
by the gravitational field, and a pair atmosphere forms. 
The probability of pair annihilation
increases because of the increase of the pair number density in
the atmosphere, and this results in decrease of the fraction of
pairs in the emerging emission in comparison with the case when
gravity of the star is neglected. 

The number rate of emerging pairs $(\dot N_\pm)$ as functions of
$\tilde L_\pm$ is shown in
Figure~4. For $\tilde L_\pm > 10^{37}$ 
ergs~s$^{-1}$ the value of
$\dot N_\pm$ is $\sim 1.5-2$ times
smaller than the same calculated by Aksenov et al. (2004) where gravity
is neglected. This is due to
partial suppression of pair creation 
as the photon energies are
reduced by gravitational redshift.
We can see that there is an upper limit to the rate of emerging
pairs $\dot N_e^{\rm max}\simeq 10^{43}$~s$^{-1}$.

Figure~5 presents the energy spectra of the emerging photons
for different values of $\tilde L_\pm$. At low luminosities,
$\tilde L_\pm\sim 10^{35}-10^{37}$ ergs~s$^{-1}$, photons that form 
in
annihilation of $e^\pm$ pairs escape from the vicinity of the
strange star more or less freely, and the photon spectra resembles
a very wide annihilation line with the mean energy of $\sim 400$~keV
(see Fig.~6). The small decrease in mean photon
energy $\langle\epsilon_\gamma\rangle$ from $\sim 430$ keV at
$\tilde L_\pm\simeq 10^{34}-10^{35}$ ergs~s$^{-1}$ to $\sim 370$ keV 
at $\tilde L_\pm\simeq
10^{37}$ ergs~s$^{-1}$ occurs because of the energy transfer from
annihilation photons to $e^\pm$ pairs via Compton scattering
(Aksenov et al. 2004, 2005).
As a result of this transfer, the emerging $e^\pm$
pairs are heated up to the mean energy $\langle\epsilon_e\rangle
\simeq 400$ keV at $\tilde L_\pm\simeq 10^{37}$ ergs~s$^{-1}$. 
For $\tilde L_\pm>
10^{37}$ ergs~s$^{-1}$, changes in the particle 
number due to
three body processes are essential, and their role in
thermalization of the outflowing plasma increases with the
increase of $\tilde L_\pm$. We see in Figure 5 that, for $\tilde 
L_\pm = 10^{42}$
ergs~s$^{-1}$, the photon spectrum is near blackbody,
except for the presence of a high-energy tail at $\epsilon_\gamma
>100$ keV. At this luminosity, the mean energy
of the emerging photons is $\sim 40$ keV, while the mean energy of
the blackbody photons is $\sim 30$ keV.

\begin{figure}
\centering
  \includegraphics[width=0.52\textwidth]{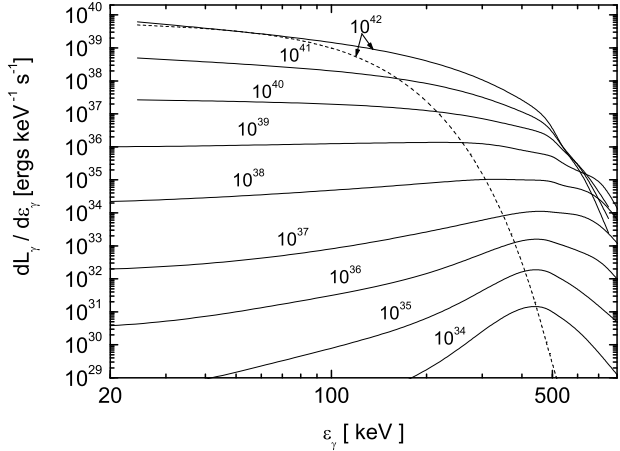}
\caption{The energy spectrum of emerging photons
for different values of $\tilde L_\pm$, as marked on the curves.
The dashed line is the spectrum of blackbody emission.}
\label{Aksenovf5.eps}       
\end{figure}

\begin{figure}
\centering
  \includegraphics[width=0.52\textwidth]{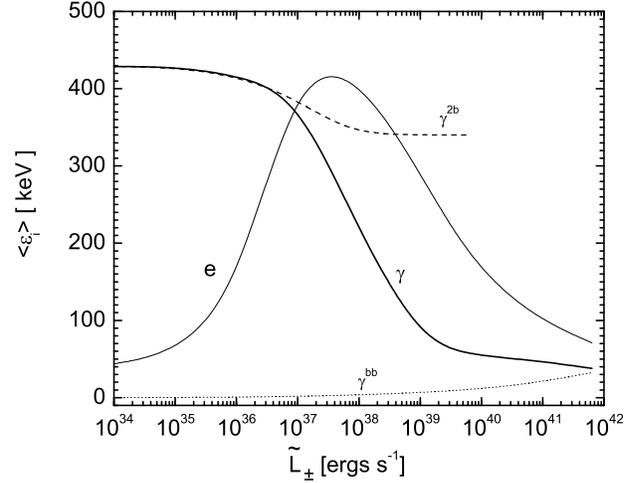}
\caption{The mean energy of the emerging photons
(thick solid line) and electrons (thin solid line) as a function
of the total injection luminosity $\tilde L_\pm$. 
For comparison, we show as the dotted
line the mean energy of blackbody photons for the same energy
density as that of the photons at the photosphere. Also shown as
the dashed line is the mean energies of the emerging photons in
the case when only two particle processes are taken into account.}
\label{Aksenovf6.eps}       
\end{figure}

Since strange quark matter at the surface of a bare strange star 
is bound via strong 
interaction rather than gravity, such a star is not subject to
the Eddington limit and can radiate in photons and pairs at the 
luminosity of $10^{51}-10^{52}$ ergs~s$^{-1}$ or even higher
(see Fig.~1). Another important idiosyncrasy that we find is 
hard spectra and a strong
anti-correlation between spectral hardness and luminosity. While
at very high luminosities ($L>10^{43}$ ergs~s$^{-1}$) the spectral
temperature increases with luminosity as in blackbody radiation,
in the range of luminosities we studied, where thermal equilibrium
is not achieved, the expected correlation is opposite (see Fig. 6).
These differ qualitatively from the photon emission from  
neutron stars and provides a definite observational signature for 
bare strange stars. 

\begin{acknowledgements}
The research was supported by the Israel Science
Foundation of the
Israel Academy of Sciences and Humanities.
\end{acknowledgements}



\end{document}